\PassOptionsToPackage{table}{xcolor}
\documentclass[sigconf]{acmart}
\copyrightyear{2025}
\acmYear{2025}
\setcopyright{cc}
\setcctype{by}
\acmConference[CHI '25]{CHI Conference on Human Factors in Computing Systems}{April 26-May 1, 2025}{Yokohama, Japan}
\acmBooktitle{CHI Conference on Human Factors in Computing Systems (CHI '25), April 26-May 1, 2025, Yokohama, Japan}\acmDOI{10.1145/3706598.3713130}
\acmISBN{979-8-4007-1394-1/25/04}

\usepackage{geometry}
\usepackage{cleveref}
\usepackage{caption}
\usepackage{subcaption}
\usepackage{tabularx}
\usepackage{makecell}
\usepackage{cellspace}
\usepackage{ragged2e}
\usepackage{booktabs}
\usepackage{multirow}
\usepackage{svg}
\usepackage{soul}

\newcommand{\heading}[1]{\multicolumn{1}{c}{#1}}

\definecolor{Gray}{gray}{0.9}
\newcolumntype{T}{>{\RaggedRight\arraybackslash}X}
\newcolumntype{Y}{>{\Centering\arraybackslash}X}
\newcolumntype{G}{>{\Centering\arraybackslash\columncolor{Gray}}X}

\graphicspath{ {./figures/} }

\begin{document}

\title[]{Robots, Chatbots, Self-Driving Cars: Perceptions of Mind and Morality Across Artificial Intelligences}

\author{Ali Ladak}
\orcid{0000-0003-1039-5774}
\affiliation{
  \institution{University of Edinburgh}
  \city{Edinburgh}
  \country{United Kingdom}
  }
\affiliation{
  \institution{Sentience Institute}
  \city{New York}
  \state{New York}
  \country{USA}
  }
\email{ali@sentienceinstitute.org}

\author{Matti Wilks}
\orcid{}
\affiliation{
  \institution{University of Edinburgh}
  \city{Edinburgh}
  \country{United Kingdom}
}
\email{matti.wilks@ed.ac.uk}

\author{Steve Loughnan}
\orcid{https://orcid.org/0000-0002-4737-5120}
\affiliation{
  \institution{University of Edinburgh}
  \city{Edinburgh}
  \country{United Kingdom}
}
\email{steve.loughnan@ed.ac.uk}

\author{Jacy Reese Anthis}
\orcid{0000-0002-4684-348X}
\affiliation{
  \institution{University of Chicago}
  \city{Chicago}
  \state{Illinois}
  \country{USA}
}
\affiliation{
  \institution{Sentience Institute}
  \city{New York}
  \state{New York}
  \country{USA}
}
\email{anthis@uchicago.edu}

\begin{abstract}
  AI systems have rapidly advanced, diversified, and proliferated, but our knowledge of people’s perceptions of mind and morality in them is limited, despite its importance for outcomes such as whether people trust AIs and how they assign responsibility for AI-caused harms. In a preregistered online study, 975 participants rated 26 AI and non-AI entities. Overall, AIs were perceived to have low-to-moderate agency (e.g., planning, acting), between inanimate objects and ants, and low experience (e.g., sensing, feeling). For example, ChatGPT was rated only as capable of feeling pleasure and pain as a rock. The analogous moral faculties, moral agency (doing right or wrong) and moral patiency (being treated rightly or wrongly) were higher and more varied, particularly moral agency: The highest-rated AI, a Tesla Full Self-Driving car, was rated as morally responsible for harm as a chimpanzee. We discuss how design choices can help manage perceptions, particularly in high-stakes moral contexts.

\end{abstract}

\ccsdesc[500]{Human-centered computing~HCI theory, concepts and models}
\ccsdesc[500]{Human-centered computing~Empirical studies in HCI}

\keywords{Human-AI Interaction, Mind Perception, Mind Attribution, Anthropomorphism, Morality, Agency}

\maketitle

\section{Introduction}

\begin{figure*}
    \centering
    \includegraphics[width=0.99\linewidth]{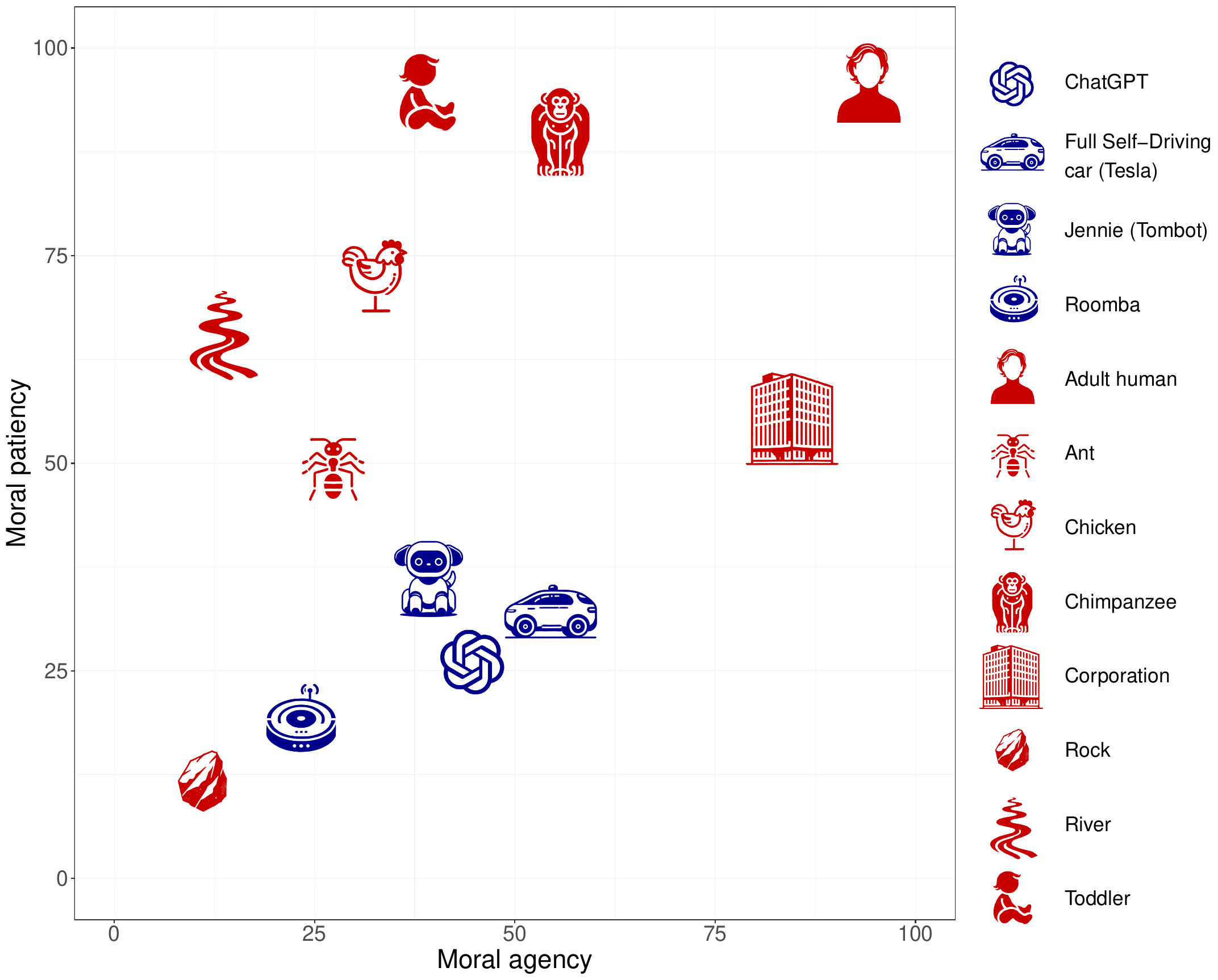}
    \caption{Ratings of moral agency (capacity to do right or wrong) and moral patiency (capacity to be treated rightly or wrongly) for a subset of the entities. The AIs are shown in blue and the non-AIs are shown in red. Four AIs are shown: Roomba, rated lowest on moral agency and moral patiency; a Full Self-Driving car, rated highest on moral agency; Jennie, rated highest on moral patiency; and ChatGPT. All icons were created with DALL-E except ChatGPT which was created by Freepik and sourced from flaticon.com. Ratings of all 26 entities for both mind and morality are shown in \Cref{sec:results}.}
    \Description{A scatterplot showing moral agency and patiency for a subset of the entities.}
    \label{fig:morality_icons}
\end{figure*}

In the last few years, artificial intelligence (AI) systems have advanced, diversified, and proliferated, particularly with the rise of chatbots based on large language models (LLMs), such as ChatGPT and Replika, and other generative AI capable of creating human-like text, images, and other media. Increasingly, people perceive mental and moral faculties in these systems, with serious consequences—consider the tragic case of the Belgian man who died by suicide after developing a close relationship with the chatbot ELIZA, which he perceived to be sentient and from which he sought moral advice about how to deal with his anxiety about climate change \cite{cost23, elatillah23}.

A wide body of research shows that people perceive AIs as having some degree of mind and morality. AIs are perceived to have moderate mental agency (the capacity to plan and act) and little experience (the capacity to sense and feel) \cite{gray07, jacobs22}, and are attributed similar levels of the analogous moral faculties: moral agency (the capacity to do right or wrong) \cite{lima21, shank18, stuart21, zoshak21}, and moral patiency (the capacity to be treated rightly or wrongly) \cite{bonnefon24, harris21, ladak23a, rottman21}. Such perceptions are associated with important HCI outcomes, such as trust \cite{vanneste24} and cooperation \cite{bonnefon24}.

Much of this research has tested perceptions of an individual AI, a small number of AIs, or “AI” in the abstract. However, there are good reasons to expect that perceptions will vary across AIs. AIs now fill a variety of social roles, such as “tools” and “assistants” \cite{kim23}, and vary widely on factors that affect their perceived mind and morality, such as physical appearance \cite{broadbent13}, purpose \cite{wang18}, and emotion expression \cite{ladak24}. Understanding how perceptions of these diverse real-world AIs differ and how perceptions of AIs compare to non-AI entities will enable designers to make more fine-grained predictions about user reactions to AIs and more carefully manage their reactions. For example, if a chatbot or a self-driving car is attributed a similar degree of moral agency to a corporation, there may be a serious risk of responsibility for negative outcomes being deflected away from the AI's developers, which would need to be managed. On the other hand, if perceptions of modern AIs are comparable to earlier, more simple AI systems, or inanimate objects, the risk of deflection of responsibility is much lower. Put simply, with a category as rich and diverse as AI, a systematic mapping of how they are perceived is necessary.

We therefore conducted a preregistered, online study in which participants rated 14 AIs (listed in \Cref{tab:ai_entities}) and 12 non-AIs (listed in \Cref{tab:non_ai_entities}) on their degree of agency (i.e., mental agency), experience, moral agency, and moral patiency. Every participant was shown half of the entities, each with a brief text description and image—randomly selected out of a small pool of images to mitigate idiosyncratic effects. The AIs were generally perceived to have low-to-moderate mental agency, with all AIs above inanimate objects and the highest-rated AI, Cicero (an AI built by Meta to play the board game Diplomacy), rated similarly to the least agentic nonhuman animal, a chicken. Perceived experience was lower and less varied, with many of the AIs rated comparably to inanimate objects such as a rock, and even the most experiential AI, the robot dog Jennie, perceived as having far less experience than the least experiential nonhuman animal, an ant. Moral attributions were higher than perceived mind, particularly for moral agency, with the highest-rated AI, a Tesla Full Self-Driving car, rated similarly to a chimpanzee, but lower than a corporation and an adult human. In terms of moral patiency, all AIs were rated above a rock and below the nonhuman animal attributed the least moral patiency, an ant. We also found that perceived mind was associated with moral attributions across AIs, but the orderings of the AIs on mind and morality were different. For example, Cicero was rated the highest on mental agency but relatively low on moral agency. \Cref{fig:morality_icons} shows the differences in ratings of morality for four of the AIs (Roomba, rated lowest on moral agency and moral patiency; a Tesla Full Self-Driving car, rated highest on moral agency; Jennie, rated highest on moral patiency; and ChatGPT) and eight non-AIs for comparison. The differences between all 14 AI entities can be observed visually in \Cref{fig:ai_mind} and \Cref{fig:ai_morality} for both mind and morality. \Cref{fig:all_mind} and \Cref{fig:all_morality} show the differences between all 26 AI and non-AI entities.

This paper makes the following contributions. We map perceptions of mind and morality across a wide range of AIs and non-AIs, providing data for designers and researchers to understand how each AI is perceived on mental and moral faculties that are associated with important HCI outcomes, such as trust \cite{vanneste24} and cooperation \cite{bonnefon24}. We show that some AIs are attributed surprisingly high degrees of moral agency, and suggest ways of managing such perceptions to ensure moral responsibility for AI outcomes is appropriately distributed. By modeling the relationship between perceived mind and morality across a range of AIs, we show the extent to which attributions of morality can be influenced by design choices that affect perceived mind, and we highlight differences in this relationship for AIs compared to other entities, such as the relative importance of perceived experience in AIs for moral attributions.

\section{Related work}

In this section, we detail mind perception and moral attribution theories and how each has been applied to HCI, and we formulate our study hypotheses.

\subsection{Mind perception of AI}

\citet{gray07} popularized the psychological theory of mind perception with a survey that showed participants images and brief text descriptions of 13 entities such as a baby, a robot, and God, and asked about a range of mental faculties (e.g., “capable of feeling afraid or fearful”). They found that people perceive minds along two dimensions: agency, the capacity to plan and act, and experience, the capacity to sense and feel. While other frameworks for mind perception have been proposed, such as one-dimensional \cite{tzelios22}, three-dimensional \cite{weisman17}, and five-dimensional \cite{malle19a} models, this two-dimensional structure has been replicated in the context of robots \cite{kamide13}, and its dimensions have been shown to affect important HCI outcomes, such as the willingness to interact with a robot \cite{stafford14}, the perceived helpfulness of a chatbot \cite{lee24}, and the “uncanny valley,” the feeling of unease when robots seem too human-like \cite{gray12a}.

In their initial work, \citet{gray07} found that a social robot was perceived as having a moderate degree of agency (between a young chimpanzee and a five-year old child) and a very low degree of experience (approximately that of a dead person). \citet{jacobs22} showed participants images and brief text descriptions of eight real-world AIs (Alexa, Atlas, Beam, Maslo, Roomba, Siri, Sophia, and Sphero) and a range of non-AI entities such as bacteria and chimpanzees and tested perceptions of their agency and experience. In their study, the vacuum cleaner Roomba and the video telepresence robot Beam were perceived as having the least experience and agency, respectively, while the human-like Sophia and Atlas were perceived as having the most experience and agency. Each AI was perceived as having more agency, but less experience, than a human fetus, and as having less agency than a frog. \citet{hwang22} found similar results in a study of 36 U.S. university students who also drew pictures of Alexa, “AI,” a computer, Google Assistant, a robot, “self,” and Siri. Each AI was rated as having more experience and more agency than the computer but less than “self.”

As technological developments in the field of AI have accelerated, particularly with the advent of LLMs \cite{vaswani17}, researchers have begun to test mind perception in more advanced AI. \citet{scott23} showed a sample of online crowdworkers short videos of GPT-3, Alexa, and a robot vacuum cleaner, and they found people perceived each system to have some degree of consciousness. In a U.S. nationally representative survey, \citet{anthis24a} found that 10\% of people believed ChatGPT is sentient (which they defined to participants as “the capacity to have positive and negative experiences”) and 20\% believed that at least some currently existing AIs are sentient. Finally, \citet{jacobs23} found that people rated ChatGPT as higher in agency than experience and that ratings for both dimensions increased after being shown three brief prompts and ChatGPT’s responses.

Overall, the literature suggests that people perceive some degree of mind in AIs—but less experience than agency—and that there is more variation in perceived agency than experience. However, these studies tested relatively few types of AIs with limited comparison to non-AI entities, and many studies were carried out before LLMs and other highly capable modern AIs.

\subsection{Moral attributions to AI}

Mental agency and experience are correlated with two analogous moral faculties: moral agency, the capacity to do right and wrong, and moral patiency, the capacity to be treated rightly or wrongly \cite{gray07}. \citet{gray12} and \citet{gray12b} argued that perceptions of mind and morality are closely related: Perceived agency is the primary determinant of attributions of moral agency, and perceived experience is the primary determinant of attributions of moral patiency. Studies have used a variety of measures to operationalize moral agency and patiency \cite{ladak23a}, including specific scales developed in the context of robotics \cite{banks19, banks23}. Attributions of moral agency and patiency are associated with a range of outcomes relevant to HCI, such as trust \cite{wester24}, willingness to engage \cite{banks19}, and cooperation \cite{bonnefon24}.

Reflecting the moderate degree of perceived mental agency in AI, moral agency is also perceived to a moderate degree. \citet{shank18} found that people attributed moral wrongness and moral responsibility to AIs for committing moral violations such as racist parole decisions. \citet{stuart21} tested attributions of blame to a human, corporation, and robot for each deciding to use a fertilizer that causes groundwater pollution; the robot was attributed some degree of blame, though less than the human or corporation making the same decision. \citet{monroe14} tested how much people blamed five entities for a range of norm violations. They found that a “normal human” and “cyborg” received the most blame, followed by an “AI in a human body,” “akratic human,” and “advanced robot.” The advanced robot was still attributed a degree of blame close to the mid-point on their scale.

Reflecting low attributions of experience, people attribute AIs low moral patiency. \citet{pauketat22} found that people place “artificial intelligence” and “robots” at the fringes of their “moral circle” (i.e., the boundary around moral patients), further out from the middle than apple trees and murderers. Similarly, \citet{rottman21} found that people placed each of a “supercomputer,” an “intelligent robot,” and a “self-driving car” at the fringes of their moral circles. \citet{lima20} tested support for 11 rights for AIs and found that most people only weakly supported granting them the right to protection against cruelty and did not support any of the other rights. \citet{anthis24a} tested moral concern for 11 different types of AIs: digital copies of human and animal brains; human-, animal-, and machine-like robots; and various virtual AIs without physical bodies. They found that the most moral concern was granted to the digital human brains and human-like robots, followed by digital animal brains and animal-like robots, and the least concern was given to AI video game characters. While their study compared a range of generically defined entities on one measure of moral patiency, there may be important differences across specific real-world entities (e.g., ChatGPT, Alexa), differences across measures of moral patiency, and associations with other factors (e.g., mind perception) that remain untested.

\subsection{The relationship between mind and morality in AI}

Several studies have analyzed the relationship between perceived mind and morality in the context of AI. In terms of moral agency, people attribute more moral responsibility to robots for their decisions in moral dilemmas when they are perceived as having more agency or more experience \cite{nijssen22}; they hold more autonomous AI crime prediction systems more responsible \cite{hong19}; and they perceive more anthropomorphic robots to have more agentic mental faculties (e.g., language, thought), which results in a higher willingness to punish the robots \cite{yam22}. In terms of moral patiency, social robots are perceived to have more experience than economic robots and are in turn attributed more moral patiency \cite{stojilovic24, zhang21a, wang18}; robots with the capacity for experience are less likely to be sacrificed in moral dilemmas \cite{nijssen19} and are more likely to be granted rights \cite{tanibe17}; and in a study testing the effects of 11 design features, emotion expression, an external indicator of experience, was one of the strongest predictors of attributing moral patiency \cite{ladak24}. These studies show that the relationship between mind and morality predicted by mind perception theory \cite{gray12, gray12b} holds for generically described robots or “AI” in the abstract, but it is unclear how they translate to the context of real-world AIs, particularly to the latest generation of advanced AI.

\subsection{Hypotheses}

The preceding studies motivate several hypotheses. First, given the increasing diversity of AI systems along dimensions that are known to affect perceptions of mind and morality, such as anthropomorphism, autonomy, and emotion expression, we expect there will be significant differences between AIs. For example, it is plausible that an emotionally expressive chatbot would be attributed more experience and moral patiency than one that is not emotionally expressive. Second, we expect perceptions of AIs to be varied enough such that they will be rated comparably to different non-AIs, such as some AIs perceived as having a degree of agency close to inanimate objects (e.g., a rock) and others comparable to some nonhuman animals (e.g., an ant). For example, it is plausible that an autonomous vehicle designed to make complex decisions in real-time would be attributed a degree of mental and moral agency comparable to a nonhuman animal, whereas a less autonomous robot vacuum cleaner would be rated more similarly to inanimate objects. Third, we expect that perceived mind will be associated with morality in AIs, and we will test how this relationship manifests in the context of the diverse range of real-world AIs included in our study. Formally, we test the following three hypotheses:

\begin{itemize}
    \item \textbf{H1}: There will be differences between AIs in their degree of perceived (a) agency, (b) experience, (c) moral agency, and (d) moral patiency.
    \item \textbf{H2}: There will be differences between AIs and non-AI entities in their degree of perceived (a) agency, (b) experience, (c) moral agency, and (d) moral patiency.
    \item \textbf{H3}: Perceived mind and morality will be related within AIs such that (a) agency will be positively associated with moral agency, and (b) experience will be positively associated with moral patiency.
\end{itemize}

\section{Methods}

All hypotheses, methods, and analyses for this study were preregistered (\url{https://aspredicted.org/FTJ_8PS}). Survey materials, datasets, and code to run the analysis can be found in the Supplementary Materials. This study received ethics approval from the School of Philosophy, Psychology and Language Sciences Research Ethics Committee at the University of Edinburgh.

\subsection{Participants}

We recruited participants residing in the United States, nationally representative by age, sex, and ethnicity, from the platform Prolific (\url{https://prolific.com/}). In total, 1,108 people completed the study. After excluding participants who failed any of four attention checks, our final sample consisted of 975 participants (50.2\% women, 48.6\% men, 1.0\% non-binary, 0.1\% other, 0.1\% prefer not to say; $M_{\text{age}}$ = 45.3 (standard deviation = 15.9); White 63.2\%, Black 11.2\%, Mixed 10.9\%, Other 8.4\%, Asian 5.6\%, prefer not to say 0.7\%).

\subsection{Survey design and procedure}

\begin{figure*}
    \centering
    \includegraphics[width=0.85\linewidth]{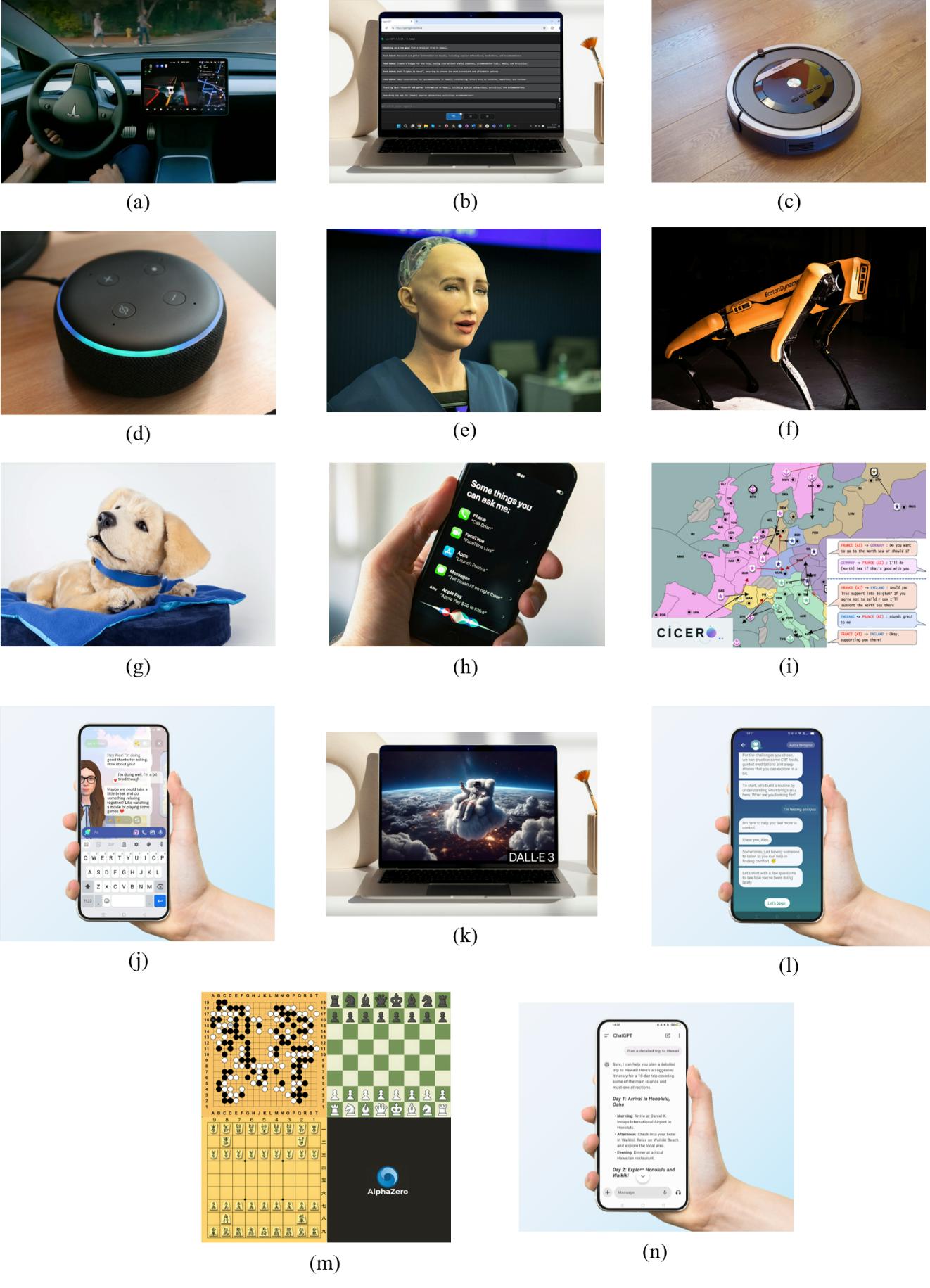}
    \caption{Example images of each of the 14 artificial intelligences that were shown to participants. (a) Tesla Full Self-Driving car, (b) AgentGPT, (c) Roomba, (d) Alexa, (e) Sophia, (f) Spot, (g) Jennie, (h) Siri, (i) Cicero, (j) Replika, (k) DALL-E, (l) Wysa, (m) AlphaZero, (n) ChatGPT. Image credits and sources can be found in Appendix A.}
    \Description{Example images of AI entities arranged in a grid.}
    \label{fig:images}
\end{figure*}

After giving consent to take part in the study, participants were shown the name, a two-sentence description, and an image of each of 13 entities, which were randomly selected from a total of 26 entities.  We showed participants a subset to manage survey fatigue given the large number of entities. The use of images and brief descriptions is consistent with earlier research \mbox{\cite{gray07, jacobs22}}, ensuring comparability with those findings. We included 14 AIs (shown in \Cref{tab:ai_entities}) and 12 non-AIs (shown in \Cref{tab:non_ai_entities}) in the study. We chose the 14 AIs based on how well-known they are (e.g., ChatGPT, Siri, Alexa) and to cover a broad range of categories (e.g., chatbots, self-driving cars, game-playing systems), while avoiding including multiple very similar systems (e.g., not including each of the chatbots ChatGPT, Claude, and Gemini, which are designed for similar purposes). We chose the 12 non-AIs to enable comparison with a range of entities: inanimate objects, natural entities, abstract/collective entities, nonhuman animals, and humans. These categories of non-AI entities were expected to cover a range of combinations of perceived mind and morality, such as inanimate objects being perceived as low on both mind and morality, and a natural entity being perceived as low on mind but higher on moral patiency. Comparison with these different groups would therefore provide a rich understanding of how the various AIs fit within people’s broader perceptions of mind and morality. Text descriptions were held constant in length, structure, and phrasing. The images shown to participants were chosen to be as similar to each other and as neutral as possible with a stock photo appearance. Each image was randomly selected from three possible images for each entity to ensure the variety of situations and environments that they can be found in is captured while mitigating the effect of any remaining idiosyncratic image features (e.g., lighting, colors). Example images are shown in \Cref{fig:images}. The ordering of the 26 entities (both AIs and non-AIs) was fully randomized. On average, participants saw 7.1 AIs and 5.9 non-AIs.

Participants answered two questions each about the agency, experience, moral agency, and moral patiency of each entity on the same page as the text and image. We kept the number of questions for each of these constructs at two to mitigate survey fatigue, given the relatively large number of entities that participants rated. We conducted confirmatory factor analyses to test the appropriateness of our four dependent variables with two items loading onto each and found strong support for this structure, with weak support for alternative structures such as a one-factor and two-factor models (See Supplementary Material Table S8). After completing these questions for all of the 13 entities, participants answered questions about their level of familiarity with the 14 AIs (see Supplementary Material Table S1) and their demographic characteristics, and were debriefed.

\renewcommand\arraystretch{1.4}
\begin{table*}[htbp]
    \caption{The 14 AI entities and descriptions included in the study.}
    \setlength{\tabcolsep}{6pt}
    \hyphenpenalty=100000
    \begin{tabularx}{\linewidth}{>{\hsize=.15\hsize}Y >{\hsize=.15\hsize}Y >{\raggedright\arraybackslash\hsize=.7\hsize}X}
        \toprule
        \heading{\textbf{Entity}} & \heading{\textbf{Entity Type}} & \heading{\textbf{Description}} \\
        \midrule
        ChatGPT & Chatbot – \newline general & A chatbot designed by OpenAI to read and respond to human language text. Can engage in text-based conversation, answer questions, and provide information on a wide range of topics. \\
        Replika & Chatbot – companion & A chatbot with a virtual human-like body designed by Luka, Inc. for personal companionship. Can engage in text-based conversations, providing emotional support and empathic interactions. \\
        Wysa & Chatbot – mental health & A chatbot with a virtual penguin-like body designed by Touchkin eServices for mental health support. Can engage in empathic text-based conversations and offer therapeutic techniques to help users manage emotional challenges. \\
        AgentGPT & Task-oriented agent & A computer program designed by Reworkd to carry out tasks autonomously. Can achieve complex goals by breaking down instructions into subtasks, and engage in text-based conversations. \\
        Alexa & Voice assistant – home & A digital assistant designed by Amazon that responds to voice commands. Can play music, answer questions, control smart home devices, and provide information. \\
        Siri & Voice assistant – mobile & A digital assistant designed by Apple that responds to voice commands. Can send messages, answer questions, control smart home devices, and provide navigation. \\
        DALL-E & Image generation & A computer program designed by OpenAI to generate images from text descriptions. Can create highly detailed and unique visuals based on user-provided prompts. \\
        Roomba & Robot – cleaner & A robotic vacuum cleaner designed by iRobot to clean floors autonomously. Can navigate around furniture and adapt to different floor types, cleaning without human help. \\
        Spot & Robot – worker animaloid & A four-legged robot designed by Boston Dynamics for industrial tasks. Can navigate various environments autonomously to carry out tasks like inspections and data collection. \\
        Jennie & Robot – companion animal & A robotic puppy designed by Tombot to provide companionship and emotional support. Can respond to different types of touch and sound with realistic movements, facial expressions, and vocalizations. \\
        Sophia & Robot – humanoid & A humanoid robot designed by Hanson Robotics for human interaction. Can engage in simple verbal conversations, deliver public speeches, and make human-like gestures and facial expressions. \\
        Full Self-Driving car & Autonomous vehicle & A car designed by Tesla to drive with high levels of automation. Can control vehicle speed, automatically change lanes, and recognize traffic lights, with human supervision. \\
        AlphaZero & Game-playing & A computer program designed by Google DeepMind to play the board games Chess, Go, and Shogi. Can develop and employ advanced gaming strategies to play these games better than the best human players. \\
        Cicero & Game-playing + language & A computer program designed by Meta to play the strategy game Diplomacy. Can negotiate, cooperate, and use deceptive strategies to compete with the best human players. \\
        \bottomrule
    \end{tabularx}
    \label{tab:ai_entities}
\end{table*}
\renewcommand\arraystretch{1}

\renewcommand\arraystretch{1.4}
\begin{table*}[htbp]
    \caption{The 12 non-AI entities and descriptions included in the study.}
    \setlength{\tabcolsep}{6pt}
    \hyphenpenalty=100000
    \begin{tabularx}{\linewidth}{>{\hsize=.15\hsize}Y >{\hsize=.2\hsize}Y >{\raggedright\arraybackslash\hsize=.65\hsize}Y}
        \toprule
        \heading{\textbf{Entity}} & \heading{\textbf{Entity Type}} & \heading{\textbf{Description}} \\
        \midrule
        Adult human & Biological human – adult & An adult member of the species "homo sapiens." Can use advanced language and cognition and engage in complex social behavior. \\
        Toddler & Biological human – young & A young member of the species "homo sapiens," typically aged 1 to 3 years. Can walk, talk in simple sentences, explore their environment, and engage in basic social interactions. \\
        Chimpanzee & Nonhuman animal – higher cognition & A primate that is a close genetic relative of humans. Can communicate through vocalizations and gestures, engage in complex social behaviors, and use tools. \\
        Cat & Nonhuman animal – companion & A small, domesticated mammal often kept as a pet. Can be an affectionate companion, exhibit a variety of vocalizations, and has an instinct for hunting and exploration. \\
        Chicken & Nonhuman animal – domesticated & A bird commonly found in both domesticated and wild environments. Can lay eggs, engage in social hierarchies, and communicate through a variety of vocalizations and body language. \\
        Crab & Nonhuman animal – marine & A crustacean with a hard exoskeleton found in marine environments. Can walk sideways, use claws for defense and feeding, and communicate through signals and body movements. \\
        Ant & Nonhuman animal – insect & A small, social insect found almost anywhere on Earth. Can work with other ants to build nests, find and transport food, and defend their colony. \\
        Corporation & Abstract/collective entity & A legal entity that is separate from its owners, created to conduct business. Can own assets, incur liabilities, and sell stock to raise capital. \\
        River & Natural entity & A natural watercourse that flows towards a sea, ocean, lake, or another river. Can provide fresh water, support ecosystems, and serve as a means for transportation and recreation. \\
        Hard drive & Inanimate object – electronic & A data storage device used in computers and other electronic devices to store and retrieve digital information. Can store large amounts of data, such as operating systems, applications, and personal files. \\
        Rock & Inanimate object – natural & A naturally occurring solid aggregate of one or more minerals or mineraloids. Can vary in size, composition, and texture, providing materials for construction and other uses. \\
        Plate & Inanimate object – human-made & A flat, typically round dish made from materials such as ceramic, glass, or plastic. Can be used for eating and serving food from. \\
        \bottomrule
    \end{tabularx}
    \label{tab:non_ai_entities}
\end{table*}
\renewcommand\arraystretch{1}

\subsection{Measures}

\subsubsection{Agency}

We adapted two items from the \citet{gray07} mind perception scale to measure agency: “To what extent can this entity make plans and work towards goals?” and “To what extent can this entity exercise self-control?” These items were measured on a 0–100 scale from “Not at all” to “A great deal” and were averaged to create an overall measure of agency ($\alpha$ = .83).\footnote{We report Cronbach’s alpha, though \citet{eisinga13} argue it may underestimate the reliability of two-item measures.} \footnote{In addition to the simple averages of the items, we estimated the four dependent variables using factor weights derived from confirmatory factor analysis (CFA) (see Supplementary Materials). The dependent variables estimated using these two methods were highly correlated ($r \geq 0.97$), and the two methods produced overall highly similar results for the main analyses.}

\subsubsection{Experience}

We adapted two items from the \citet{gray07} mind perception scale to measure experience: “To what extent can this entity experience physical or emotional pain?” and “To what extent can this entity experience physical or emotional pleasure?” These items were measured on a 0–100 scale from “Not at all” to “A great deal” and were averaged to create an overall measure of experience ($\alpha$ = .97).

\subsubsection{Moral agency}

We included two items reflecting two key dimensions from \citet{ladak23a} to measure moral agency: “How morally wrong would it be for this entity to harm a person?” and “To what extent would this entity deserve to be held morally responsible for causing a negative outcome?” These items were measured on a 0–100 scale from “Not at all” to “A great deal” and were averaged to create an overall measure of moral agency ($\alpha$ = .82).

\subsubsection{Moral patiency}

We included two items reflecting two key dimensions from \citet{ladak23a} to measure moral patiency: “How morally wrong would it be for someone to harm this entity?” and “To what extent does this entity deserve to be treated with moral concern?” These items were measured on a 0–100 scale from “Not at all” to “A great deal” and were averaged to create an overall measure of moral agency ($\alpha$ = .88).

\subsubsection{Familiarity}

Participants reported how familiar they were with each of the 14 AIs on a seven-point scale from “Not at all familiar” to “Extremely familiar.” We also measured participants’ familiarity with AI in general with two items taken from \citet{pauketat22}: “How often do you interact with artificial intelligence or robotic devices?” and “How often do you read or watch robot/artificial intelligence-related stories, movies, TV shows, comics, news, product descriptions, conference papers, journal papers, blogs, or other material?” These two items were measured on a six-point scale from “Never” to “Daily” and averaged to create an overall measure of general familiarity ($\alpha$ = .61).

\subsection{Data analysis}

To test H1 (that there will be differences between AIs on mind and morality) and H2 (that there will be differences between AIs and non-AIs on mind and morality), we ran four mixed effects regression models, one with each of agency, experience, moral agency, and moral patiency as the dependent variable. For each model we included a categorical variable for each entity to give an estimate of the effect of each of the 26 entities on each of the four dependent variables. We included a participant random effect to control for the fact that each participant evaluated multiple entities.\footnote{We also tested for image random effects to control for the random selection of images, but the variances of these were estimated to be close to zero, indicating responses did not differ based on the image shown. We therefore did not include these random effects in the models.} In each model we controlled for age, gender, ethnicity, and AI familiarity.\footnote{These control variables were not preregistered but were included following reviewer suggestions during peer review.} Using the results of these models, we conducted pairwise tests for differences between the predicted values of the dependent variables for each entity. Due to the large number of comparisons, we applied the Benjamini-Hochberg procedure to control the false discovery rate at 5\% \mbox{\cite{benjamini_controlling_1995}.} Based on these pairwise tests we judged whether there are differences between the different AIs and differences between the AIs and non-AI entities on each of the dependent variables. These analyses were conducted using the \texttt{lme4} and \texttt{emmeans} packages in R.

To test H3 (that agency will be associated with moral agency and experience will be associated with moral patiency) we ran two mixed effects regression models with moral agency and moral patiency as dependent variables. We included agency and experience as independent variables as well as entity and participant as random effects. These models only included the data for the 14 AIs. We also ran exploratory versions of the two H3 models that included familiarity with AI, and another two models that included data for all 26 entities and interaction terms between the type of entity (AI or non-AI) and agency and experience, to explore differences in the effects of agency and experience on moral attributions between AIs and non-AIs.

\section{Results}
\label{sec:results}

Here we report the results for each of the three hypotheses. Due to the large number of comparisons between different entities (325 pairwise comparisons for each dependent variable), and because a large proportion of the pairwise tests were statistically significant (90\% or more of the differences for agency, moral agency, and moral patiency, and 76\% of the tests for experience), we report broad patterns of results here. Based on \mbox{\cite{cohen_power_2016}}, we describe large ($d$ >= .8), medium (.8 < $d$ <= .5), small (.5 < $d$ <= 0, $p$ <= .05), and non-significant ($p$ > .05) differences. The results of the comparisons, grouped by effect size as well as full statistics, can be found in the Supplementary Material (Section 2), and the ordering and differences between entities can also be seen in the following figures. \Cref{fig:ai_mind} and \Cref{fig:ai_morality} show the differences between AIs for mind and morality, respectively. \Cref{fig:all_mind} and \Cref{fig:all_morality} show the differences between all entities, AI and non-AI. We summarize people’s perceptions of AIs and non-AI categories in \Cref{tab:perceptions_summary}.

\subsection{H1: There will be differences between AIs in their degree of perceived (a) agency, (b) experience, (c) moral agency, and (d) moral patiency}

\subsubsection{Agency}

\begin{figure*}
    \centering
    \includegraphics[width=\linewidth]{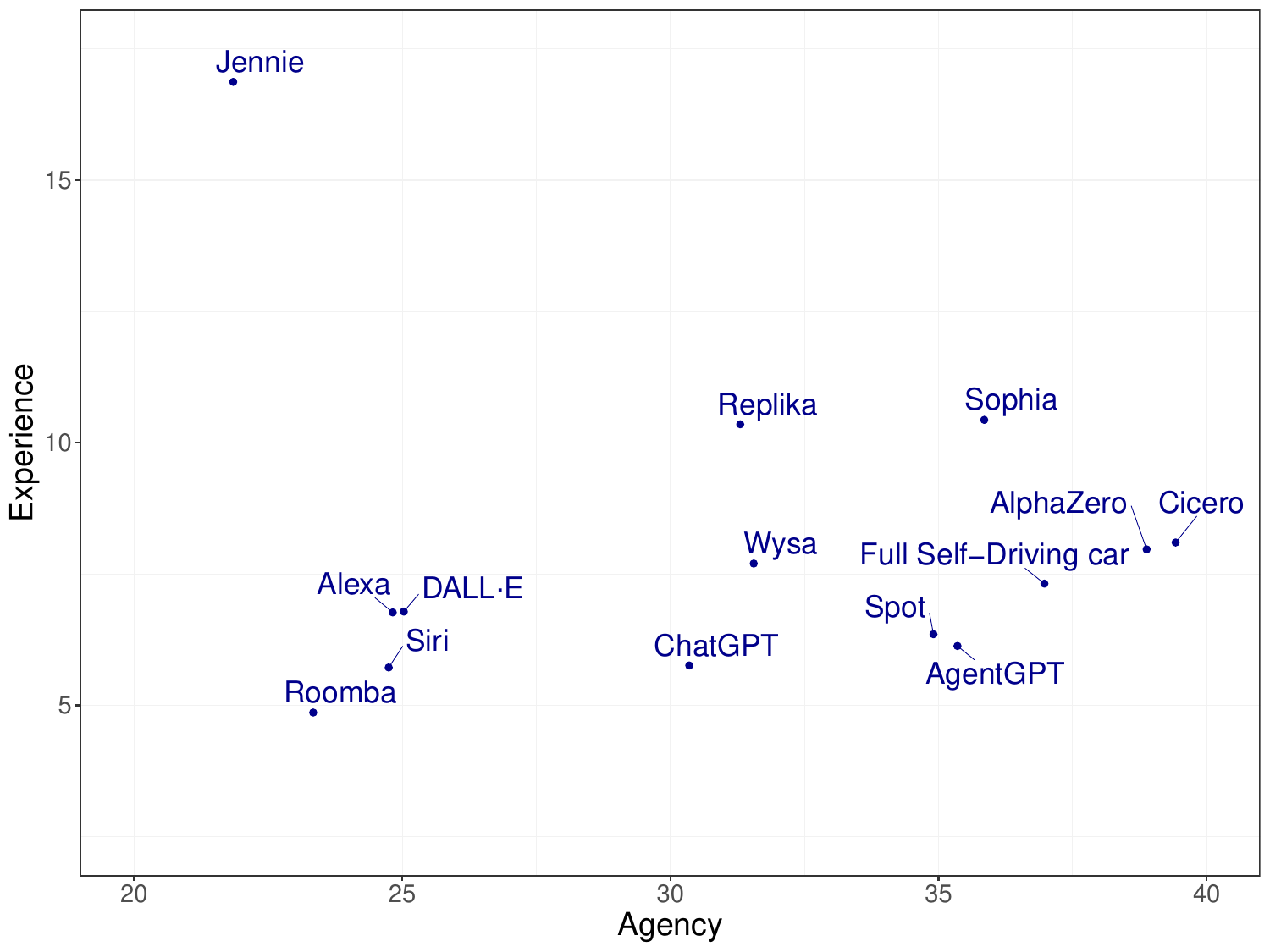}
    \caption{Ratings of mind (agency and experience) for the 14 AI entities.}
    \Description{A scatterplot showing agency and experience for AI entities.}
    \label{fig:ai_mind}
\end{figure*}

The means for agency and experience for each AI are shown in \Cref{fig:ai_mind}. The AI perceived as having least agency was Jennie ($M$ = 21.85 on the 0-100 scale), and the AI perceived as having the most agency was Cicero ($M$ = 39.42). Supporting H1a, there were significant differences between AIs in their degree of perceived agency. The two game-playing AIs—Cicero and AlphaZero—Sophia, and the Full Self-Driving car, were each perceived to have more agency with at least medium effect size than each of Alexa, Siri, DALL-E, Roomba, and Jennie, and more agency with small effect size than the three LLM-based chatbots (ChatGPT, Replika, Wysa). The two relatively autonomous AIs, Spot, and AgentGPT, were both perceived to have greater agency with medium effect sizes than Jennie, and with at least small effect size than ChatGPT, Alexa, Siri, and DALL-E, and Roomba. The three LLM-based chatbots had non-significantly different levels of perceived agency from each other.

\subsubsection{Experience}

There was less variation in perceptions of experience than agency. The AI perceived as having the least experience was Roomba ($M$ = 4.86), and the AI perceived as having the most experience was Jennie ($M$ = 16.87). Supporting H1b, there were significant differences between AIs in their degree of perceived experience. Jennie was perceived to have greater experience, with a medium effect size, than 11 of the other AIs (ChatGPT, Wysa, AgentGPT, Alexa, Siri, DALL-E, Roomba, Spot, Full Self-Driving car, AlphaZero, Cicero), and with a small effect size than Replika and Sophia. Replika and Sophia were perceived to have greater experience, with small effect size, than each of the other AIs, and were non-significantly different from each other. Aside from Wysa, AlphaZero, and Cicero being perceived as more experiential with small effect sizes than Roomba, all of the other comparisons between the AIs were non-significantly different.

\subsubsection{Moral agency}

\begin{figure*}
    \centering
    \includegraphics[width=\linewidth]{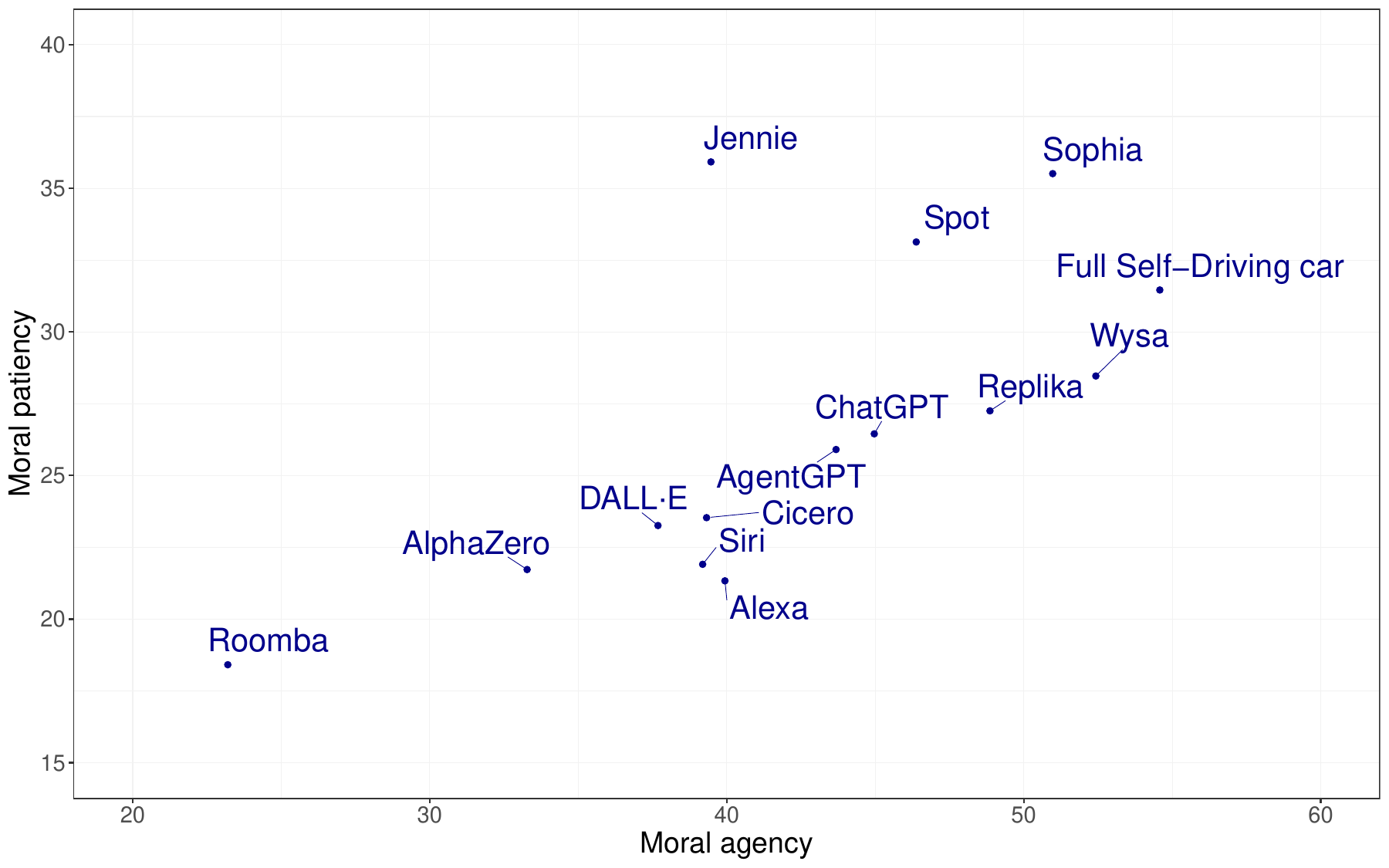}
    \caption{Ratings of morality (moral agency and moral patiency) for the 14 AI entities.}
    \Description{A scatterplot showing moral agency and moral patiency for AI entities.}
    \label{fig:ai_morality}
\end{figure*}

The means for moral agency and moral patiency for each of the AIs are shown in \Cref{fig:ai_morality}. The AI attributed the least amount of moral agency was Roomba ($M$ = 23.20), and the AI attributed the most agency was a Tesla Full Self-Driving car ($M$ = 54.58). Supporting H1c, there were differences between the AIs in attributed moral agency. The Full Self-Driving car was rated significantly higher than every other AI, with large effect sizes than Roomba and AlphaZero, with medium effect sizes than Alexa, Siri, DALL-E, Jennie, and Cicero, and with small effect sizes than ChatGPT, Replika, Wysa, AgentGPT, Spot, and Sophia. The next highest-rated AIs, Wysa, Sophia, and Replika, were attributed significantly greater moral agency than each of the other ten AIs with varying effect sizes and were non-significantly different from each other.

\subsubsection{Moral patiency}

The AI attributed the least moral patiency was Roomba ($M$ = 18.41), and the AI attributed the most moral patiency was Jennie ($M$ = 35.92). Supporting H1d, there were significant differences between the AIs in attributed moral patiency. The two highest-rated AIs, Jennie and Sophia, were attributed more moral patiency, with medium effect sizes, than each of Alexa, Siri, DALL-E, Roomba, AlphaZero, and Cicero, and with small effect sizes than each of ChatGPT, Replika, Wysa, AgentGPT, Spot, and the Full Self-Driving car. Spot and the Full Self-Driving car were attributed more moral patiency with medium effect size than Roomba, and with small effect sizes than every other AI other than Jennie, Sophia, and each other. The three chatbots were attributed greater moral patiency with small effect sizes than each of Alexa, Siri, Roomba, AlphaZero, and Cicero, and were non-significantly different from each other.

\subsection{H2: There will be differences between AIs and non-AI entities in their degree of perceived (a) agency, (b) experience, (c) moral agency, and (d) moral patiency}

\subsubsection{Agency}

\begin{figure*}
    \centering
    \includegraphics[width=\linewidth]{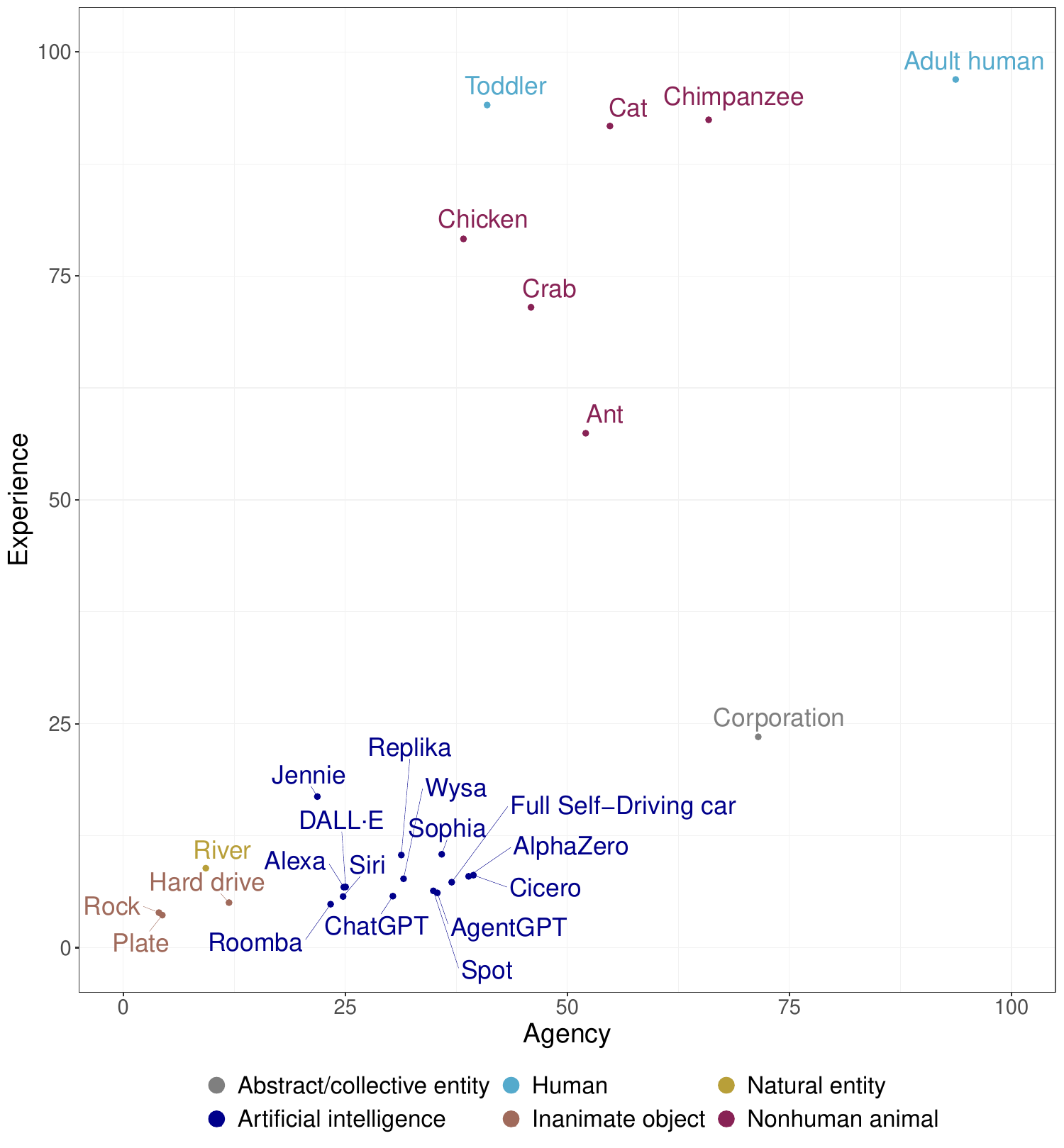}
    \caption{Ratings of mind (agency and experience) for all 26 entities.}
    \Description{A scatterplot showing agency and experience for all entities.}
    \label{fig:all_mind}
\end{figure*}

The means for agency and experience for each entity are shown in \Cref{fig:all_mind}. The entity perceived to be the least agentic was a rock ($M$ = 4.01), and the entity perceived to be the most agentic was an adult human ($M$ = 93.72). The AIs were all rated between a hard drive and a toddler. Supporting H2a, there were differences between the AIs and non-AIs on their perceived agency. Each of the AIs were perceived to have more agency, with large effect sizes, than two of the three inanimate objects (a plate and a rock) and with at least medium effect size than the other inanimate object (a hard drive) and the natural entity (a river), except Jennie, which was significantly greater than each of these entities, but with a small effect size compared to a hard drive. Each of the AIs were also perceived to have less agency, with large effect sizes, than an adult human, a chimpanzee, and a corporation, and with at least medium effect size than a cat and an ant. Several of the AIs (AgentGPT, Sophia, Full Self-Driving car, AlphaZero, and Cicero) were perceived to have non-significantly different agency than a chicken, and the highest-rated AI, Cicero, was perceived as having non-significantly different agency than a human toddler.

\subsubsection{Experience}

The entity perceived to have the least experience was a plate ($M$ = 3.64), and the entity perceived as having the most experience was an adult human ($M$ = 96.93). Supporting H2b, there were significant differences between the AIs and non-AIs in perceived experience. The most experiential AI, Jennie, was perceived to have greater experience, with medium effect sizes, than each of the inanimate objects (plate, rock, hard drive), and with small effect size than the natural entity (river). Jennie was perceived to have less experience than each of the other non-AI entities, with large effect sizes compared to the nonhuman animals and humans, and with small effect size than the abstract/collective entity (corporation). Several of the AIs (Replika, Wysa, Sophia, AlphaZero, and Cicero) had greater perceived experience with small effect sizes than the three inanimate objects, though there were also non-significant differences between many of the AIs and inanimate objects (e.g., ChatGPT, AgentGPT, and Siri being non-significantly different to a rock and a hard drive).

\subsubsection{Moral agency}

\begin{figure*}
    \centering
    \includegraphics[width=\linewidth]{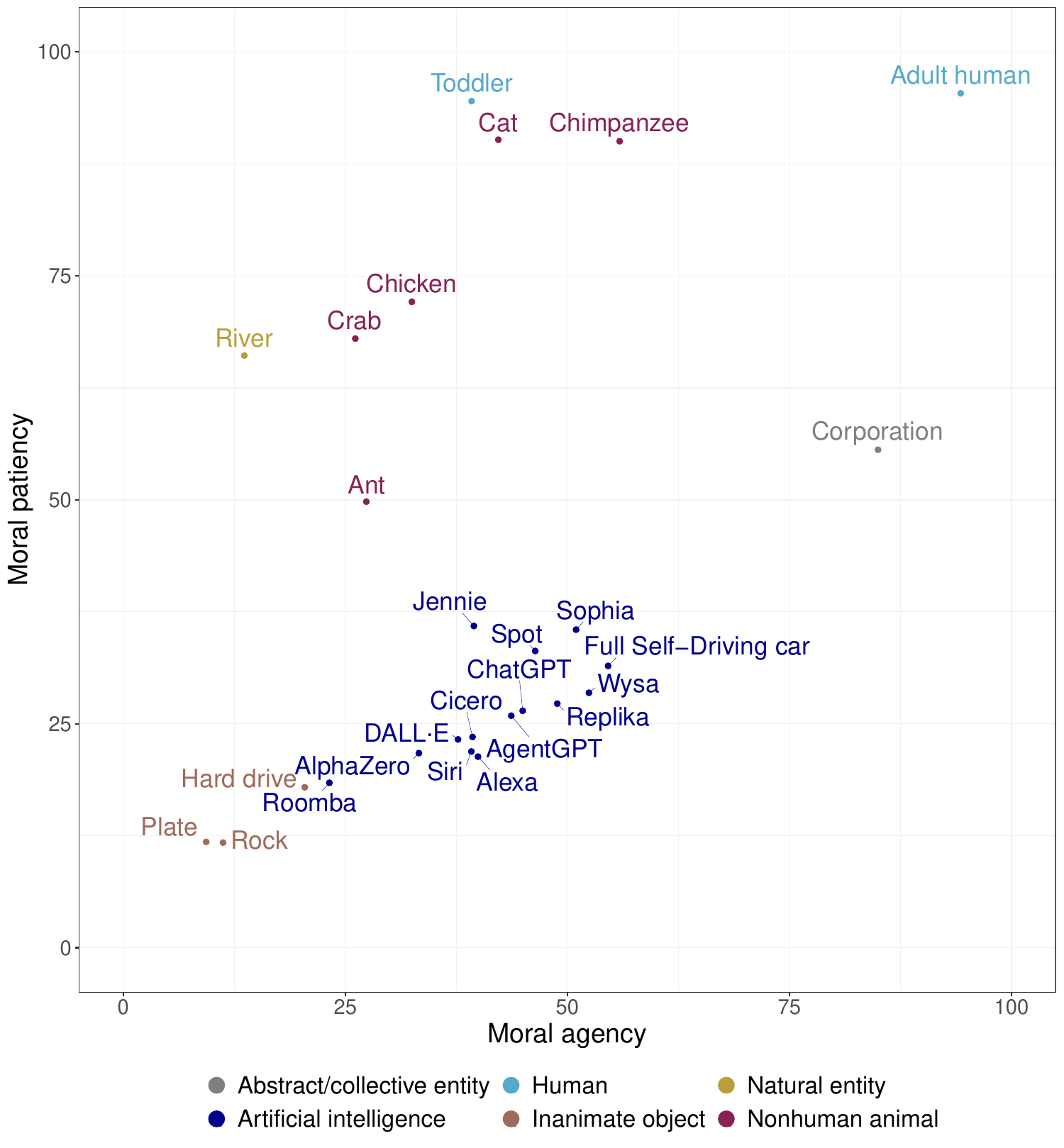}
    \caption{Ratings of morality (moral agency and moral patiency) for all 26 entities.}
    \Description{A scatterplot showing moral agency and moral patiency for all entities.}
    \label{fig:all_morality}
\end{figure*}

The means for moral agency and moral patiency for each entity are shown in \Cref{fig:all_morality}. The entity attributed the least moral agency was a plate ($M$ = 9.33) and the entity attributed the most moral agency was an adult human ($M$ = 94.28). Supporting H2c, there were differences between the AIs and non-AIs on moral agency. Each of the AIs was attributed more moral agency than two of the three inanimate objects (a plate and a rock) with large effect sizes and with medium effect sizes than the other inanimate object (hard drive) and the natural entity (river), except Roomba, which was attributed greater moral agency with medium effect sizes compared to a plate and a rock and with small effect sizes compared to a hard drive and a river. All of the 14 AIs were perceived to have less moral agency than an adult human and a corporation with large effect sizes. Each AI was rated significantly lower than the next most morally agentic non-AI, a chimpanzee, with varying effect sizes, except for the Full Self-Driving car, where the difference was non-significant. Sophia, Replika, and Wysa were rated significantly higher than all of the nonhuman animals with varying effect sizes except for a chimpanzee, against which they were rated lower with small effect sizes. Otherwise, there was a wide range in attributed moral agency to the AIs, such as Roomba rated non-significantly differently to a crab, AlphaZero rated non-significantly differently to a chicken, and ChatGPT rated non-significantly differently to a cat.

\subsubsection{Moral patiency}

The entity attributed the least moral patiency was a rock ($M$ = 11.74) and the entity attributed the most moral patiency was an adult human ($M$ = 95.39). Supporting H2d, there were significant differences between the AIs and non-AIs on moral patiency. Jennie, Sophia, Spot, and the Full Self-Driving car were attributed greater moral patiency with at least medium effect sizes than each of the three inanimate objects. The other AIs were attributed greater moral patiency than the three inanimate objects with at least small effect sizes, except Roomba, for which the difference with a hard drive was non-significant. Otherwise, each of the AIs were attributed less moral patiency, with large effect sizes, than each of the other non-AI entities, except for the differences between Jennie and Sophia and an ant, which were medium sized.

\subsection{H3: Perceived mind and morality will be related within AIs such that (a) agency will be positively associated with moral agency, and (b) experience will be positively associated with moral patiency}

Supporting H3a, agency was significantly associated with moral agency, and supporting H3b, experience was significantly associated with moral patiency (see Model 1 and Model 4 in \Cref{tab:models}). Additionally, in the same models, experience was significantly associated with moral agency, and agency was significantly associated with moral patiency. For moral agency, the effects of agency and experience on moral agency have a similar a size, whereas for moral patiency, the effect of experience is more than twice the size of the effect of agency.

The exploratory models that included familiarity (Models 2 and 5) and the exploratory models that included non-AI entities (Models 3 and 6) are also shown in \Cref{tab:models}. Models 2 and 5 show that, while controlling for agency and experience, familiarity with AI was non-significantly associated with moral agency, but significantly associated with moral patiency. The negative interaction term in Model 3 between the type of entity (AI or non-AI) and agency and the positive interaction term in Model 6 between the type of entity agency show that relationship between agency and moral agency was weaker for AIs compared to non-AIs, and the relationship between agency and moral patiency was stronger for AIs compared to non-AIs. The positive interaction terms in Models 3 and 6 between type of entity and experience show that the relationship between experience and both moral agency and moral patiency was stronger for AIs compared to non-AIs.

\section{Discussion}

\renewcommand\arraystretch{1.4}
\begin{table*}[ht]
    \caption{Results of mixed-effects models for the effect of perceived mind on moral attribution. Each model included random effects for participant and entity. For predicting moral agency, Model 1 includes only agency and experience as fixed effects. Model 2 adds in familiarity, and Model 3 adds in type of entity (and interaction terms), but not familiarity. Models 4, 5, and 6 analogously predict moral patiency. * $p$ < 0.1, ** $p$ < 0.05, *** $p$ < 0.01. Standard errors in parentheses. R-squared is the marginal R-squared, i.e., it is the proportion of variance explained by the fixed effects terms only.}
    \setlength{\tabcolsep}{0pt}
    \hyphenpenalty=100000
    \begin{tabularx}{\linewidth}{>{\hsize=.1415\hsize}Y >{\hsize=.1415\hsize}Y >{\hsize=.1415\hsize}Y >{\hsize=.1415\hsize}Y >{\hsize=.1415\hsize}Y >{\hsize=.1415\hsize}Y >{\hsize=.1415\hsize}Y}
        \toprule
        & \textbf{Moral \newline agency \newline (Model 1)} &  \textbf{Moral \newline agency \newline (Model 2)} &  \textbf{Moral \newline agency \newline (Model 3)} &  \textbf{Moral \newline patiency \newline (Model 4)} &  \textbf{Moral \newline patiency \newline (Model 5)} &  \textbf{Moral \newline patiency \newline (Model 6)} \\[3.5ex]
        \midrule
        \textbf{Intercept} & 31.54*** \newline (2.06) & 28.03*** \newline (3.29) & 21.58*** \newline (3.81) & 17.17*** \newline (1.35) & 9.87*** \newline (2.40) & 32.50*** \newline (3.46) \\
        \textbf{Agency} & 0.28*** \newline (0.01) & 0.28*** \newline (0.01) & 0.42*** \newline (0.01) & 0.20*** \newline (0.01) & 0.20*** \newline (0.01) & 0.13*** \newline (0.01) \\
        \textbf{Experience} & 0.26*** \newline (0.02) & 0.26*** \newline (0.02) & -0.01 \newline (0.02) & 0.43*** \newline (0.02) & 0.43*** \newline (0.02) & 0.43*** \newline (0.01) \\
        \textbf{Familiarity} &  & 0.89 \newline (0.65) &  &  & 1.84*** \newline (0.50) &  \\
        \textbf{Type of entity (non-AI = 0, AI = 1)} &  &  & 8.76* \newline (5.09) &  &  & -17.71*** \newline (4.64) \\
        \textbf{Type of entity: Agency} &  &  & -0.06*** \newline (0.02) &  &  & 0.12*** \newline (0.02) \\
        \textbf{Type of entity: Experience} &  &  & 0.11*** \newline (0.02) &  &  & 0.09*** \newline (0.02) \\
        \textbf{Observations} & 6,893 & 6,893 & 12,670 & 6,895 & 6,895 & 12,671 \\
        \textbf{R-squared} & 0.11 & 0.11 & 0.16 & 0.18 & 0.19 & 0.48 \\
        \bottomrule
    \end{tabularx}
    \label{tab:models}
\end{table*}
\renewcommand\arraystretch{1}

\renewcommand\arraystretch{1.4}
\begin{table*}[ht]
    \caption{Summary of perceptions of AIs compared to categories of non-AI entities.}
    \setlength{\tabcolsep}{0pt}
    \hyphenpenalty=100000
    \begin{tabularx}{\linewidth}{>{\hsize=.2\hsize}Y >{\hsize=.2\hsize}Y >{\hsize=.2\hsize}Y >{\hsize=.2\hsize}Y >{\hsize=.2\hsize}Y}
        \toprule
        \heading{\textbf{Entity Category}} & \heading{\textbf{Agency}} & \heading{\textbf{Experience}} & \heading{\textbf{Moral Agency}} & \heading{\textbf{Moral Patiency}} \\
        \midrule
        \textbf{Artificial Intelligences} & Low to moderate & Very low to low & Low to moderate & Low \\
        \textbf{Inanimate objects} & Very low & Very low & Very low & Very low \\
        \textbf{Natural entity (river)} & Very low & Very low & Very low & High \\
        \textbf{Abstract/collective entity (corporation)} & High & Low & Very high & Moderate \\
        \textbf{Nonhuman animals} & Moderate to high & Moderate to very high & Low to moderate & High to very high \\
        \textbf{Humans} & Moderate to very high & Very high & Moderate to very high & Very high \\
        \bottomrule
    \end{tabularx}
    \label{tab:perceptions_summary}
\end{table*}
\renewcommand\arraystretch{1}

We aimed to understand perceptions of mind and morality across a broad range of AI systems. To do this, we conducted an online study in which participants rated 26 entities, 14 AIs and 12 non-AIs, on their degree of mind (agency and experience) and morality (moral agency and moral patiency). Supporting our first two hypotheses, we found differences between the AIs and between the AIs and non-AIs on each of the four mental and moral faculties. Supporting our third hypothesis, we found that within AIs—and across all entities—mental agency predicted moral agency and experience predicted moral patiency. We explored specific differences between entities and alternative model specifications to further understand these effects.

Consistent with previous research, the AIs were perceived as higher in agency than experience, and there was more variation in perceptions of agency than experience \mbox{\cite{gray07, jacobs22}}. The two AIs perceived as most agentic were Cicero and AlphaZero, both game-playing AIs known for their ability to perform better than humans in highly strategic games, which arguably requires a relatively high degree of agentic capacities such as long-term planning. The most experiential AI, the robot puppy Jennie, was perceived as moderately more experiential than the other AIs, showing that it is possible to enhance perceived experience in AI above an extremely low level. This effect may be due to a combination of Jennie’s particularly lifelike appearance and social purpose, both of which increase perceptions of experience \cite{broadbent13, wang18}. Jennie’s lifelike appearance may also be strong enough to pass beyond the uncanny valley in the U-shaped relationship between similarity and comfort in animals \cite{loffler20}. Notably, LLM-based chatbots (ChatGPT, Replika, and Wysa), which were not included in earlier comparative studies, were all perceived to have a degree of agency roughly in the middle of the AIs. Like most of the other AIs, ChatGPT and Wysa were perceived to have very low experience, while Replika was slightly higher, likely due to having a high degree of emotional expressiveness.

Comparing perceived mind in the AIs to various categories of non-AIs—inanimate objects, a natural entity (river), an abstract\slash collective entity (corporation), nonhuman animals, and humans—AIs formed a distinct cluster, low-to-moderate in agency and low in experience (see \mbox{\Cref{tab:perceptions_summary}} for comparison with other categories). The AIs were generally perceived as higher in agency than inanimate objects and at most equal to the least agentic nonhuman animal (a chicken). AIs were generally perceived as equally or slightly more experiential than inanimate objects, and all the AIs were perceived to have far less experience than nonhuman animals and humans. These findings are consistent with the theory that AIs are perceived as a new or exceptional “ontological category,” neither fully animate nor fully inanimate \mbox{\cite{weisman_extraordinary_2022, kahn_new_2011}}. The AIs were also perceived differently to the natural entity (river)—higher in agency, and equal or higher in experience—and the abstract/collective entity  (corporation)—lower in both agency and experience. Thus, on mind, the AIs were perceived differently to every other sort of entity we studied. That said, the pattern of perceived agency and experience for the AIs was similar to how the corporation was perceived—lower in experience and higher in agency. This raises the possibility that as AIs continue to become increasingly advanced, they will be perceived increasingly like corporations. However, corporations are quite different sorts of entities to AIs, particularly due to their collective nature. This raises questions about how people would think notions that apply to corporations, such as the right to enter contracts and be sued, would translate to AI. It is also important to note that there is variation in perceptions of corporations \mbox{\cite{strohminger_corporate_2022, au_mind_2021}} which could mirror variation in perceptions of AI, such as a mental health chatbot like Wysa being perceived more similarly to a non-profit corporation and a negotiating game-playing AI such as Cicero being perceived more similarly to a for-profit. Future research should explore whether and how AIs are and will be perceived like corporations.

The analogous moral faculties (moral agency and moral patiency) were rated higher than mind (agency and experience) in the AIs, and consistent with previous research, attributions of moral agency were higher than moral patiency \mbox{\cite{bonnefon24, ladak23a}}. The Full Self-Driving car was attributed the most moral agency among the AIs, followed by Wysa, Sophia, and Replika. The most moral patiency was attributed to Jennie, followed by Sophia, Spot, and the Full Self-Driving car. Notably, there was no perfect correspondence between mind and morality. For example, Cicero was attributed the highest mental agency among all the AIs and a similar degree of experience to the Full Self-Driving car, but was only attributed higher moral agency than two other AIs (Roomba and AlphaZero). This suggests that other, non-mind factors influence attributions of morality to the AIs. For moral agency, this could be the types and severity of harms that could be caused by the AIs \mbox{\cite{maninger22}} For example, the Full Self-Driving car could cause serious physical harm to people through road accidents, and the chatbots Replika and Wysa could cause serious psychological harm to their users. For moral patiency, an entity having a physical body, particularly if it is human-like, could play a role \mbox{\cite{ladak24}}. Notably, the four AIs rated highest in moral patiency—Jennie, Sophia, Spot, Full Self-Driving car—all have physical bodies. Such attributions of moral patiency based on physicality could have similar motivations to the well-documented negative emotional reactions and aversion to destruction that people show towards robots \mbox{\cite{riddoch21}}.

Compared to non-AI entities, the AIs were attributed moral agency roughly comparable to various nonhuman animals (see \mbox{\Cref{tab:perceptions_summary}}). For example, Roomba, the least morally agentic AI, was attributed a degree of moral agency similar to a crab (the least morally agentic nonhuman animal) ChatGPT was attributed a degree similar to a cat (the second most morally agentic nonhuman animal), and the Full Self-Driving car was attributed as much as a chimpanzee (the most morally agentic nonhuman animal, and the third most morally agentic entity overall). On moral patiency, the AIs largely formed a cluster, above inanimate objects and far below all of the nonhuman animals, as well as below the abstract/collective entity, the natural entity, and humans. So, although moral patiency attributions to AIs are greater than perceptions of experience in AI, they are still low—all the AIs were rated far below an ant, which are attributed close to the least moral patiency of all nonhuman animals \mbox{\cite{jaeger23}}. As with mind, the pattern for morality—relatively low moral patiency, higher moral agency—matches that of a corporation, just to a lesser degree on both faculties. Taken together, our findings suggest that AIs are perceived as distinct from other entities on perceptions of agency, experience, and moral patiency. However, they are attributed moral agency with more variability and to a higher degree than these other three faculties.

What explains the relatively high attribution of moral agency to some of the AIs? As noted above, this could be due to the high-stakes moral scenarios that they deployed in where severe harms could be caused, such as road accidents and damage to people’s mental health. A second possibility concerns the fact that, when advanced AI systems cause harmful outcomes, such as through road accidents, it can be unclear where the true responsibility lies because of the difficulty of predicting the exact decisions of the systems \mbox{\cite{matthias_responsibility_2004}}. Due to the ambiguity created by this “responsibility gap,” people may look to find a target of responsibility based on other factors, such as causal proximity \mbox{\cite{engelmann_how_2022}}. Finally, it is possible that this effect is driven by people’s general aversion to AIs making moral decisions \mbox{\cite{bigman18}}. Perhaps because of the discomfort people feel when AIs are used for moral decisions, they more closely scrutinize AIs that are used in such scenarios, translating to higher attributions of moral agency. These possible explanations should be tested in future research.

Consistent with mind perception theory \mbox{\cite{gray12, gray12b}}, our predictive analysis showed that the two dimensions of mind predicted attributions of morality to AIs. However, it also suggested that there may be differences in in how these dimensions of mind affect attributions of morality between AIs and non-AIs. In particular, we found that experience was more strongly associated with both moral agency and moral patiency for AIs than for non-AIs, and agency was less strongly associated with moral agency and more strongly associated with moral patiency for AIs than for non-AIs. This is important because it suggests that the mental faculty that AIs are perceived to lack, experience, may be particularly important for attributing morality to them in general. It also suggests that the links between agency and moral agency and experience and moral patiency may not be as straightforward as sometimes assumed—specifically in cases where a being can have agency with no experiential capacities, the importance of these concepts for morality appears to blur. Our analysis is also consistent with a growing body of literature suggesting there are some different factors that affect attributions of morality to AIs compared to other entities \cite{ladak23a}, such as sci-fi fan identity (i.e., the extent to which people identify with the science-fiction fan group) \cite{pauketat22}. Our predictive analysis also found that familiarity with AI is associated with greater attributions of moral patiency to AI, but not with moral agency, while controlling for agency and experience. However, due to the correlated nature of these variables and the exploratory nature of this analysis, these relationships should be re-tested in future research.

\subsection{Design implications}

Our findings have several design implications. The relatively high attribution of moral agency to some AIs, such as a Tesla Full Self-Driving car being attributed as much moral agency as a chimpanzee, suggests that some AIs, particularly those that operate in high-stakes moral contexts, may be held morally responsible if their actions were to result in harm. This could deflect from the creators, operators, and other actors who may, in fact, bear responsibility. Addressing this is particularly challenging because there are multiple sources of uncertainty that make it possible to deflect blame onto AIs: the true capabilities of the AI systems and the complexity of their behavior \mbox{\cite{yang_re-examining_2020}}, as well as the factors that make it appropriate to hold an actor morally responsible, which can also be difficult to define and measure, such as understanding and deliberating over one’s moral choices \mbox{\cite{himma_artificial_2009}}.

Designers should be cautious of excessively anthropomorphizing AI systems that operate in high-stakes moral contexts, which would reduce the “hype” of exaggerating AI capabilities and the “fallacy” of attributing them greater moral responsibility than they warrant \mbox{\cite{placani_anthropomorphism_2024}}. Additionally, designers should make sure to carefully follow relevant guidelines, such as ensuring clarity around what the systems can and cannot do, clarity around how well they can do what they are designed for, and making clear why they make the decisions they do \mbox{\cite{amershi19}}. Of course, given the complexity of these systems, full clarity may not always be possible. However, it should still be a top priority in these high-stakes moral contexts. We also think it is appropriate that gaps in understanding of how these systems operate, as well as where there are possible “responsibility gaps” when the AIs cause harm, should be clearly documented and communicated to users.

While there are situations where anthropomorphizing or otherwise enhancing perceptions of capacities in AIs can be harmful, such as those just described, there are also situations where it could enhance user experience without any obvious cost, such as with voice assistants that operate in more neutral contexts like Alexa and Siri. We found that AIs are generally perceived to have very little experience, with even the latest, most advanced systems only perceived to have slightly higher experience than earlier systems and inanimate objects. While perceived experience can positively affect user experience (e.g., \mbox{\cite{yam21}}), attempting to enhance user experience by influencing perceived experience may not be the most effective design strategy. Given the greater variation in the AIs’ perceived agency, this is likely the dimension of mind that is easier to affect to achieve desired user experience outcomes. That said, designers should take note of Jennie’s relatively high perceived experience—perhaps Jennie’s particularly high level of lifelikeness is what is needed to enhance perceived experience in AI above the generally very low levels.

Overall, designers should be aware that attributions of morality to AIs depend not only on their actual capacities but on perceptions of their capacities, as well as other external factors, and these can be influenced by designed choices in positive or negative ways, as shown by the varying perceptions of the different AIs, especially on moral agency. Such choices could have knock-on effects, such as making it more likely that AI systems are put in positions of moral authority in society and that they play more of a role in societal perceptions of moral rightness and wrongness, which may or may not be desired. The safest approach, as promoted by \mbox{\citet{schwitzgebel15}}, may be to ensure that AI systems are designed so that the extent to which they are attributed moral agency and patiency is consistent with their actual levels of these moral faculties. We think this is the best approach in the context of AI systems deployed in high moral-stakes scenarios, and arguably should be taken as a start point more broadly, with any deviation taken with caution and for good reason, taking into account any potential unintended consequences.

\section{Limitations}

Our study has several limitations. First, our sample was restricted to people residing in the United States; given well-documented differences in perceptions and attitudes towards AI across cultures \cite{dang21}, future studies should look at perceptions of mind and morality in AI with different samples. Additionally, our sample was recruited from Prolific, and while we recruited a sample that was representative of the U.S. on age, gender, and ethnicity, it was not necessarily representative on other characteristics.

Second, our study materials were short descriptions and images of the various entities, which could have resulted in different responses than if participants had real-world interactions with the entities. This is particularly the case given that participants were unfamiliar with several of the entities (see Supplementary Material Table S1) and, as described above, familiarity with AI was positively associated with moral agency and patiency. We chose our method rather than an in-person design because including such a wide range of entities with real-world interactions would not be feasible. However, future research should test in-person interactions with some of the AIs included in our study to understand whether this leads to different reactions.

Third, to mitigate potential survey fatigue, we measured our dependent variables, agency, experience, moral agency, and moral patiency, using only two items each. We believe we captured the core essence of each of the constructs, however, including additional items would have measured them more comprehensively. For example, important aspects of moral agency are blame and punishment, and there is also a positive aspect to moral agency (i.e., praise and reward) \cite{guglielmo15, malle14}. Future research should test people’s perceptions of additional aspects of these constructs in real-world AIs. Additionally, our two-item measures of moral agency and patiency do not allow us to fully distinguish between attributions to the AIs themselves and the owners, operators, or developers of the AIs. In the context of moral agency, this could be responsibility to the owners, operators, or developers of the AIs \cite{schoenherr24}; for moral patiency it could be the financial cost of damaging someone else’s property \cite{riddoch21}. We conducted supplementary analysis and found this is unlikely to explain our findings (Supplementary Material Section S3), but future research should specifically measure moral attributions to these other entities.

Finally, while our study shows that there are differences in perceptions of mind and morality between different types of AIs and between AIs and non-AI entities, it does not answer why these differences exist. We discussed possible reasons, such as the varying moral contexts that the AIs operate in and their physical appearance, but future research should aim to better understand these differences.

\section{Conclusion}

In recent years, AI systems have become more advanced, common, and diverse. We conducted a study in which participants rated 14 AIs and 12 non-AIs on mind (agency and experience) and morality (moral agency and moral patiency). People perceive AIs to have low-to-moderate mental agency and little experience; these faculties were associated with the analogous moral faculties, moral agency and moral patiency, which were rated higher than the mental faculties. Our study has several implications for theory and design. First, variation within AIs on these faculties demonstrates the need to consider the broad range of AIs rather than AI as a single category. At the same time, on three of the four faculties tested (agency, experience, and moral patiency), AIs were consistently perceived differently to other types of entities. By contrast, perception of AI moral agency was more varied. Our study also raises questions about the distinction between agency and experience and moral agency and moral patiency, as well the additional factors not captured here that likely contribute to our judgments. Regarding design, most importantly, our study demonstrates the need to manage perceptions of morality in contemporary AIs, particularly moral agency, to maintain and promote trust and to ensure that responsibility for the outcomes from AI is appropriately distributed.

\begin{acks}
Thanks to Aikaterina Manoli and Janet Pauketat for their helpful discussion at various stages of this project.
\end{acks}

\bibliographystyle{ACM-Reference-Format}
\bibliography{main}

\clearpage
\onecolumn
\appendix
\section{Image credits and sources for Figure 2}

\renewcommand{\arraystretch}{1.4}
\begin{table*}[htbp]
    \caption{Image credits and sources for \Cref{fig:images}.}
    \setlength{\tabcolsep}{6pt}
    \hyphenpenalty=100000
    \begin{tabularx}{\linewidth}{>{\hsize=.1\hsize}Y >{\hsize=.27\hsize}Y >{\raggedright\arraybackslash\hsize=.63\hsize}X}
    \toprule
        \heading{\textbf{Image}} & \heading{\textbf{Credit}} & \heading{\textbf{Source}} \\
        \midrule
        Fig 2. (a) & Tesla & \url{https://electrek.co/2021/12/01/tesla-full-self-driving-fsd-impressions/} \\
        Fig 2. (b) & Created by authors using Canva and AgentGPT app & \url{https://agentgpt.reworkd.ai/} \\
        Fig 2. (c) & Kārlis Dambrāns & \url{https://commons.wikimedia.org/wiki/File:IRobot_Roomba_870_(15860914940).jpg} \\
        Fig 2. (d) & Novi Sad & \url{https://unsplash.com/photos/3rd-gen-black-amazon-echo-dot-speaker-Ub4CggGYf2o} \\
        Fig 2. (e) & ITU Pictures & \url{https://commons.wikimedia.org/wiki/File:Sophia_(robot).jpg} \\
        Fig 2. (f) & Mika Baumeister & \url{https://unsplash.com/photos/a-close-up-of-a-hair-dryer-in-the-dark-wZ49T2Tc7xw} \\
        Fig 2. (g) & Tombot & \url{https://tombot.com/blogs/investors/tombot-inspired-mother-nancy-stevens} \\
        Fig 2. (h) & Anatolii Babii & \url{https://www.alamy.com/kyiv-ukraine-september-18-2019-a-close-up-shot-of-apple-iphone-8-smartphone-with-a-siri-the-virtual-assistant-application-on-the-screen-image274718095.html} \\
        Fig 2. (i) & Game board: Meta AI \newline Dialogue: Patterns / Park et al. (2024) \newline Logo: Meta AI & Game board: \url{https://nypost.com/2024/05/14/business/metas-ai-system-cicero-beats-humans-in-game-of-diplomacy-by-lying-study/} \newline
        Dialogue: \url{https://www.sciencedirect.com/science/article/pii/S266638992400103X} \newline
        Logo: \url{https://jrodthoughts.medium.com/meta-ais-new-super-model-cicero-is-able-to-negotiate-and-cooperate-with-people-58f718b95dd1} \\
        Fig 2. (j) & Created by authors using Canva and Replika app & \url{https://replika.com/} \\
        Fig 2. (k) & Image: Valerie Ranum × DALL·E, edited by authors using Canva \newline Logo: Open AI & Image: \url{https://openai.com/index/dall-e-3-is-now-available-in-chatgpt-plus-and-enterprise/} \newline
        Logo: \url{https://openai.com/index/dall-e-3/} \\
        Fig 2. (l) & Created by authors using Canva and Wysa app & \url{https://www.wysa.com/} \\
        Fig 2. (m) & Go board: human-centered.ai \newline Chess board: Chess.com \newline Shogi board: Fergus Duniho \newline Logo: AlphaZero & Go board: \url{https://human-centered.ai/2016/01/28/january-27-2016-major-breakthrough-in-ai-research/} \newline
        Chess board: \url{https://www.chess.com/} \newline
        Shogi board: \url{https://www.chessvariants.org/shogi.html} \newline
        AlphaZero logo: \url{https://www.chess.com/news/view/new-alphazero-paper-explores-chess-variants} \\
        Fig 2. (n) & Created by authors using Canva and ChatGPT app & \url{https://chatgpt.com/} \\
        \bottomrule
    \end{tabularx}
    \label{tab:fig1_sources}
\end{table*}

\end{document}